\documentclass[aps,prl,preprint,groupaddress]{revtex4}

\usepackage{graphicx}
\usepackage{amsmath}
\usepackage{amsfonts}
\usepackage{amssymb}
\usepackage{dcolumn}
\usepackage{bm}

\newcommand{\beqn}{\begin{eqnarray}}
\newcommand{\eeqn}{\end{eqnarray}}
\newcommand{\be}{\begin{equation}}
\newcommand{\ee}{\end{equation}}

\def\beq{\begin{equation}}
\def\eeq{\end{equation}}
\def\beqn{\begin{eqnarray}}
\def\eeqn{\end{eqnarray}}

\def\du{d_{\cal U}}
\def\mstar{M_*}
\def\mpl{M_{pl}}
\def\lu{\Lambda_U}
\def\pd{\frac{1}{\du -1}}

\def\s1{$s_{\alpha}$}
\def\s2{$s_{\gamma}$}
\def\s3{$s_{\delta}$}
\def\c1{$c_{\alpha}$}
\def\c2{$c_{\gamma}$}
\def\c3{$c_{\delta}$}
\def\s{Stueckelberg~}

\newcommand{\mathsym}[1]{{}}

\begin{document}

\title{Scalar modifications to gravity from unparticle effects may be  testable}

\author{  Haim Goldberg and   Pran Nath}

\affiliation{Department of Physics, Northeastern University,
Boston, MA 02115, USA }

\begin{abstract}
Interest has focussed  recently on low energy implications of  a
nontrivial scale invariant sector of an effective field theory
with an IR fixed point, manifest in terms of ``unparticles'' with
peculiar properties. If unparticle stuff exists it could couple to
the stress tensor and mediate a new    'fifth' force which we call 
'ungravity' arising from the exchange of unparticles
between massive particles, which in turn could modify the inverse square
law. Under the assumption of strict  conformal  invariance in the
hidden sector  down to low energies, we compute the lowest order
ungravity correction to the Newtonian gravitational potential and
find scale invariant power law corrections of type
$(R_{G}/r)^{2d_{\cal U} -1}$ where  $d_{\cal U}$  is an anomalous
unparticle dimension and $R_{G}$ is a characteristic  length scale
where the ungravity interactions become  significant. $d_{\cal U}$
is constrained to lie the range  $ d_{\cal U} > 3 (2)$ for a spin
2 (spin 0) unparticle coupling  to the stress tensor (and its
trace) and  leads to modification of the inverse square law
with $r$  dependence in the range between  $1/r^{4+2\delta} (\delta>0)$,
while extra dimension models with warping modify the force law
with corrections beginning with terms O$(1/r^3)$  for small $r$
but exponentially suppressed for large $r$.
Thus a  discrimination between extra dimension models and ungravity is possible in future
improved submillimeter tests of gravity.

\end{abstract}
\maketitle
Recently much interest has been generated by the possibility that a  nontrivial scale
invariant sector of an effective field theory\cite{Banks:1982gt}
characterized by unparticles\cite{Georgi:2007ek} could play a
role in low energy physics. This has led
to several further investigations of unparticle effects in collider physics  and elsewhere
~\cite{Cheung:2007ue}.
In  models of this type one assumes that the ultraviolet theory has hidden sector operators  $O_{UV}$ of dim $d_{UV}$
possessing  an IR fixed point. These
couple via  a connector sector with the Standard Model operators $O^n_{SM}$
of dimension $n$, generating an effective interaction
${O_{UV} O^n_{SM}}/{M_U^{d_{UV}+n-4}}$.
It is then assumed that the fields of the hidden sector undergo
dimensional transmutation at scale $\Lambda_U$
generating scale
invariant unparticle  fields $O$ with dimension $d_{\cal U}$ which give the interaction
\beqn
\left(\frac{\Lambda_U}{M_U}\right)^{d_{UV}+n-4} \frac{O O^n_{SM}}{\Lambda_U^{d_u+n-4}} 
\eeqn
The operator $O$ could be a scalar, a vector, a tensor or even a spinor. If $O$ is a
tensor of rank two
it could  couple  to the stress tensor and its exchange
between physical particles could lead to a modification of Newtonian gravity. We discuss
this issue in this note.

We work strictly within the framework  where  conformal invariance
holds down to low energies, and thus we forbid scalar unparticle operators  of
dimension $d_{\cal U} <2$
which could have couplings  to the Higgs  field of the type $H^2O$.
The presence of  a super-renormalizable operator destroys conformal
invariance once $H$ develops a VEV~\cite{Foxetal}.
   In our analysis here  we consider an effective operator of the type
\beqn
\kappa_* \frac{1}{\Lambda_U^{d_u-1} }\sqrt g  T^{\mu\nu}  O^{\cal U}_{\mu\nu},
\label{int}
\eeqn
where $\kappa_*$ is defined by   $\kappa_*= \Lambda_U^{-1} (\Lambda_U/M_U)^{d_{UV}} $.
We assume that $O_{\mu\nu}^{\cal U}$ transforms like a tensor under the
general co-ordinate transformations,
and thus the interaction of Eq.(\ref{int}) gives an action which is
invariant under the transformations.
For convenience we assume that $O_{\mu\nu}^{\cal U}$   is traceless.
The addition of Eq.(\ref{int})  to the action
changes  the stress-energy tensor so that the new  tensor is
$
{\cal T}^{\mu\nu} = T^{\mu\nu} +   ({\kappa_*}/{\Lambda_U^{d_u-1} })
g^{\mu\nu}
T^{\sigma \rho} O^{\cal U}_{\sigma\rho}$.
The conservation condition in this case is
$
{\cal T}^{\mu\nu}_{;\nu}=0.
$
The interaction of Eq.(\ref{int}) implies
that the  unparticles  can be exchanged between massive particles  and this 
exchange creates a new force, a 'fifth' force, which we call 'ungravity' which
adds to the force of gravity. 
We wish to compute the correction to the
Newtonian gravitational potential arising from the exchange of the unparticles to the lowest order.
In this case
one may neglect all the gravitational effects and replace   $g_{\mu\nu}$ by
$\eta_{\mu\nu}$.
The quantity that  enters in the computation of the unparticle exchange contribution is the
unparticle propagator
\beqn
\Delta_{\cal U}^{\mu\nu\sigma\rho}(P) =\int e^{iPx} <0|T(O_{\cal U}^{\mu\nu}(x) O_{\cal U}^{\sigma\rho})|0> d^4x.
\eeqn
An analysis  of this propagator using spectral
decomposition~\cite{Georgi:2007ek,Cheung:2007ue}  gives
\beqn
\Delta_{\cal U}^{\mu\nu\sigma\rho}(P) =  \frac{A_{d_{\cal U}} }{\sin(\pi d_{\cal U})}
  P^{\mu\nu\sigma\rho} (-P^2)^{d_{\cal U}- 2},
\label{propagator}
\eeqn
where  $P^{\mu\nu\sigma\rho}$ has the form
$
P^{\mu\nu\sigma\rho}  = \frac{1}{2} (P^{\mu\sigma}P^{\nu\rho} + P^{\mu\rho} P^{\nu\sigma}
- \alpha P^{\mu\nu} P^{\sigma\rho}).
$
Here $\alpha=2/3$ and
$P^{\mu\nu}=(-\eta^{\mu\nu}+ P^{\mu} P^{\nu}/P^2)$,  and $A_{d_{\cal U}}$ is given by
\be
A_d = \frac{16\pi^{5/2}}{(2\pi)^{2\du}}\ \frac{\Gamma(\du + 1/2)}{\Gamma(\du - 1)\Gamma(2\du)}
\label{adu}
\ee
Further, since we  are  interested in  computing the effects of the unparticles  to the lowest
order,   we  can replace  ${\cal T}^{\mu\nu}$ by ${ T}^{\mu\nu}$
and  replace   $ {\cal T}^{\mu\nu}_{;\nu}=0$
by ${ T}^{\mu\nu}_{,\nu}=0$.  This condition implies that  momentum factors when acting on the
source give  a vanishing contribution and the relevant part of $P^{\mu\nu\sigma\rho}$ can be written as
$
P^{\mu\nu\sigma\rho} = \frac{1}{2} ( \eta^{\mu\sigma}\eta^{\nu\rho} +\eta^{\mu\rho}\eta^{\nu\sigma}
-\alpha\eta^{\mu\nu}\eta^{\sigma\rho}).
$
For the case of the graviton exchange $\alpha=1$ and retaining $\alpha$ in the analysis will provide
a quick check to the graviton exchange  limit.

We have carried out an analysis of  the unparticle exchange  along with the one graviton exchange
and computed the effective potential in the non-relativistic limit.  
We find
\beqn
V(r)= -m_1m_2 G    \left( \frac{1}{r} + 
\frac{\xi^2}{\Lambda_U^{2d_{\cal U}-2} }
\frac{ (2-\alpha)}{ (2\pi)^{2(d_{{\cal U} } -1) } }
\frac{2}{ \sqrt {\pi}}
\frac{\Gamma (2-d_{\cal U}) \Gamma (d_{\cal U}+ \frac{1}{2})}   { \Gamma (2d_{\cal U})}f_{d_{\cal U}}(r)\right),
\label{N2}
\eeqn
where $\xi = \kappa_*/\kappa$  and $\kappa= M_{Pl}^{-1}$  where $M_{Pl}$ is the Planck mass
$M_{Pl}= 2.4\times 10^{18}$ GeV, and
where $f_{d_{\cal U}}(r) $ is given by
$
f_{d_{\cal U}}(r) =4\pi \int ({d^3 q}/{(2\pi)^3})   {e^{-i{\bf q.r} }}/ {({\bf q}^2)^{2-d_{\cal U}}}.
$
The first term in the brace in Eq.(\ref{N2}) is the Newtonian term, while the second
term is the ungravity correction.  
One can easily check that the ungravity correction produces the correct Newtonian  potential
for  the case  $d_{\cal U} =1$ and  $\alpha=1$ since $f_{d_{\cal U}}(r) =1/r$ in this case.
However,  for $d_{\cal U}$ different from unity the ungravity effects produce an  $r$ dependence
of the form $1/r^{2d_u-1}$ which  can be  differentiated from the  effects
of  ordinary gravitation. An  explicit evaluation gives
\beqn
V(r)= -\frac{m_1m_2 G}{r}  \left[ 1
+\left(\frac{R_{G}}{r}\right )^{2d_{\cal U}-2}\right ],\nonumber\\
R_{G}=  \frac{1}{\pi\Lambda_U} \left(\frac{M_{\rm Pl}}{M_*}\right)^{\frac{1}{d_{\cal U}-1}}.
\left(
\frac{2(2-\alpha)} {\pi}
\times \frac{\Gamma (d_{\cal U}+ \frac{1}{2}) \Gamma (d_{\cal U}- \frac{1}{2})
}   { \Gamma (2d_{\cal U})}\right)^{\frac{1}{2d_{\cal U}-2}}
\label{RG}
\eeqn
 where  $M_*=\kappa_*^{-1}$. The choice 
$d_{\cal U} < 1$ will produce corrections to the gravitational potential which fall off slower than
$1/r$,  and thus would modify the very large  distance  behavior of the gravitational potential,  which
appears  undesirable.   Thus a sensible constraint on  $d_{\cal U}$ is $d_{\cal U} >1$
in which case the ungravity contribution to
the potential falls off faster than $1/r$ and the short distance behavior will be affected.
Constraints of conformal invariance in this case  require\cite{mack,nakayama}
$d_{\cal U}>1+s$ where $s$ is the spin of the operator, and  thus for a rank one tensor
operator $d_{\cal U}>2$ and  for a  rank two $d_{\cal U}>3$. Let us now consider a spin zero
unparticle operator with $d_{\cal U} \geq 2$ with coupling of the type
$\kappa_* \sqrt g  T^{\mu}_{;\mu}  O^{\cal U}/\Lambda_U^{d_{\cal U}-1}$.
In this case the modified Newtonian potential can be gotten from Eq.(\ref{RG}) by
replacing $(2-\alpha)$ by 2.  
With this choice 
the  corrections to the potential can begin with terms
of $O(1/r^{(4+2\delta)}), \delta>0$.
\\

\begin{figure*}[t]
\centering
\hspace*{.1in}\includegraphics[width=7.2cm,height=6.4cm]{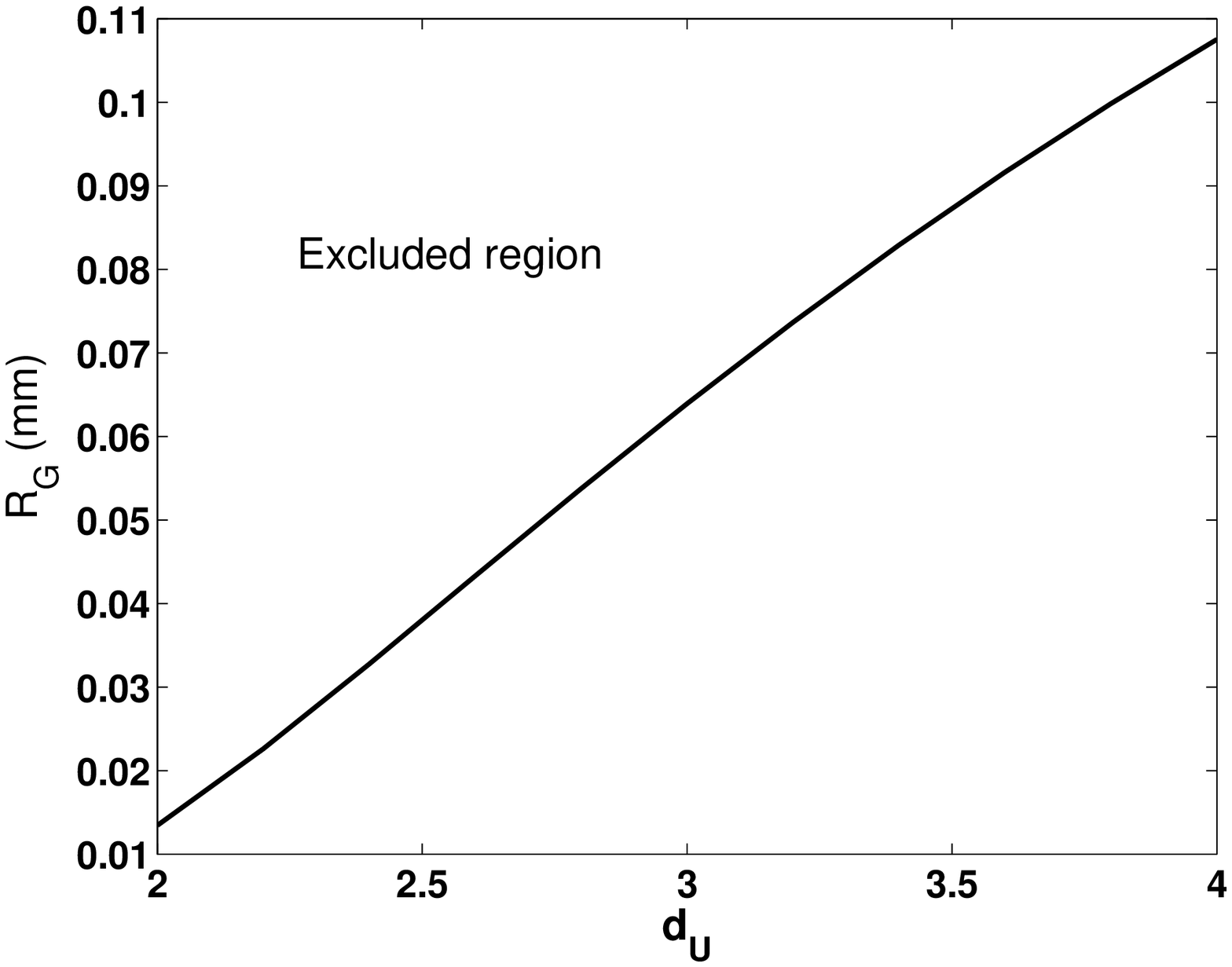}
\hspace*{0.2in}\includegraphics[width=7.8cm,height=6.4cm]{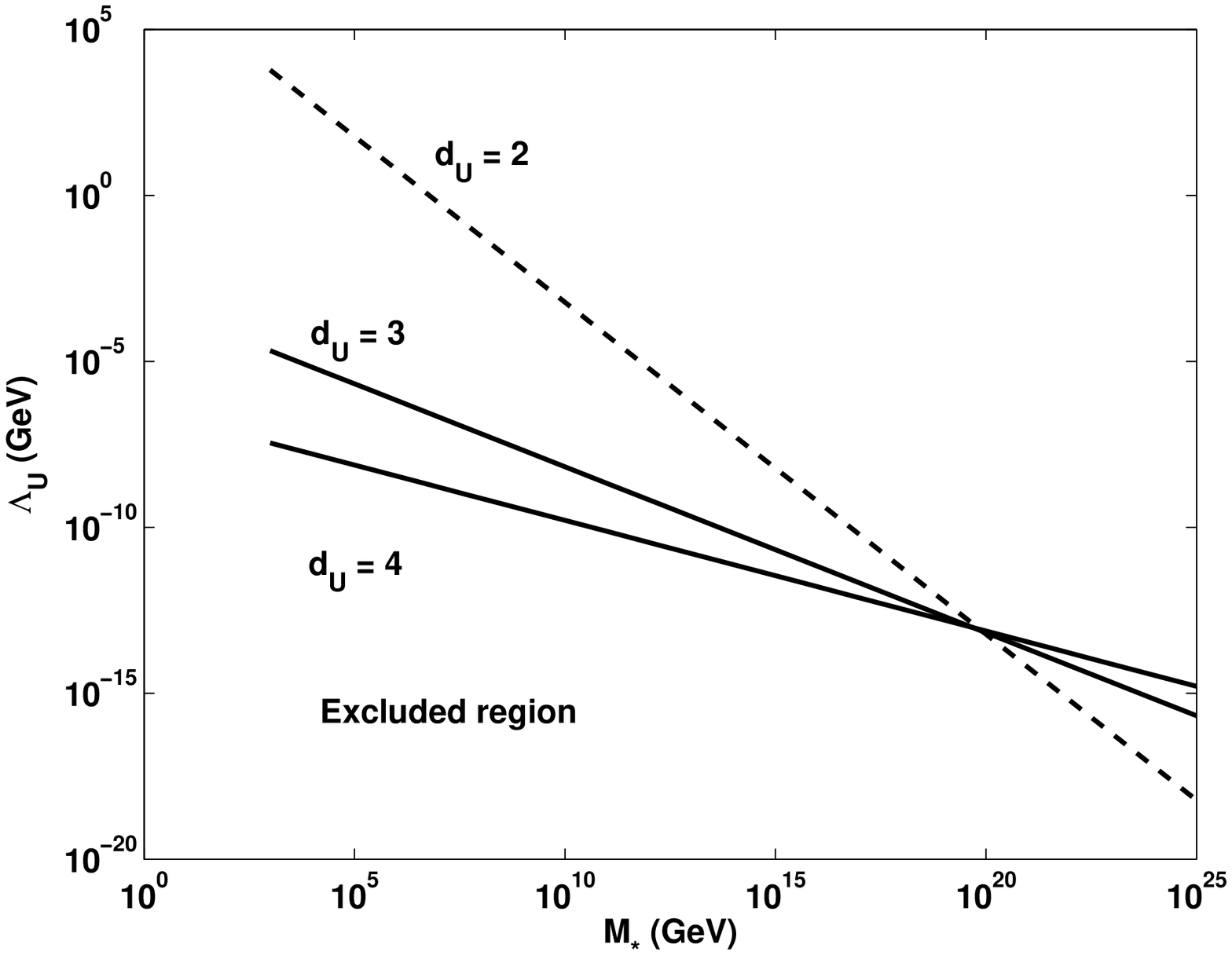}
\caption{Left side: allowed region (below curve) for $R_G$
(Eq.~(\ref{RG})) for a region of $d_{\cal U}$. Right side: Allowed
region in $M_*-\Lambda_{\cal U}$ parameter space (above curves)
for various values of $d_{\cal U}$. The seeming confluence of the
three lines at a single point is not exact.} \label{fig12}
\end{figure*}

   The short distance ungravity contribution is
constrained by the recent  precision submillimeter  tests of the gravitational inverse square law
\cite{Kapner:2006si,Adelberger:2006dh}.
The current experiment probes short distances up to around .05 mm, and no significant deviation from the inverse
square law is seen.
 However, better  precision experiments
in the future  will be able to explore the parameter space of the model more thoroughly.
Returning now to Eq.~(\ref{RG}), the quantity $R_G$ may be
constrained by experiment. The usual parameterization of the
corrrection to Newtonian gravity in terms of a Yukawa term is not
directly suitable for the present case. Instead, we extrapolate
the power law limits in Table I of Ref.~\cite{Adelberger:2006dh}
to obtain an upper limit on $R_G$ as a function of $d_{\cal U}.$
The result of this exercise is shown in the left panel of
Fig.(\ref{fig12}), where the current data excludes the region
above the curve. In the right panel in Fig.(\ref{fig12}) we
present an analysis of the allowed region of the $\Lambda_{\cal
U}- M_*$ parameter space which follows from Eq.~(\ref{RG}) when
combined with the constraint on $R_G$. Here the regions below the
curves are excluded by the current data.

It is of interest that for $M_*\simeq M_{Pl}$ the  value of
$\Lambda_{\cal U}$ required for proximity to the present bound is
very low. In order to assess this possibility and  explore the
constraint of a conformal fixed point we
   examine an $SU(N)$ gauge theory with $N_f$
 massless Dirac fermions.  In this case an infrared fixed point occurs at a coupling
 \cite{Appelquist:1996dq}
  $
 \alpha_*= -4\pi(11N- 2N_f)/$ $(34N^2-$ $10NN_f$ $-3(N^2-1)N_f/N)
 $.
For values of $N_f$ close to and below $11N/2$ but above $N_f^c
=N(100N^2-66)/(25N^2-15)$ where the  chiral symmetry breaking
occurs, one is in the region of a conformal fixed point. In this
region
the scale  $\Lambda_{\cal U}$ is roughly given by the scale $\Lambda$ in
Ref.~\cite{Appelquist:1996dq}:
   \beqn
   \Lambda_{\cal U} \approx M_G \exp\left[-\frac{1}{b\alpha*} \ln\left(\frac{\alpha_*-\alpha(M_G)}{\alpha(M_G)}\right)
   -\frac{1}{b\alpha(M_G)}\right]
   \eeqn
   where $b=(11N-2N_f)/ 6\pi$.
   Thus for $N=3$, the region of the conformal fixed point is $16.5>N_f>11.9$.
   To get an estimate we  set  $M_G=1\times 10^{16}$GeV,
   $\alpha(M_G)\simeq 0.04$, $N=3$, $N_f=12$,  and find an infrared fixed point at
   $\alpha_*= 0.75$  which gives $\Lambda_{\cal U}\approx 10^{-11}$ GeV.
   This is an explicit demonstration that  an IR fixed point can occur with  $\Lambda_{\cal U}$
   very small, which is of interest in our analysis.

The modification of gravity discussed here differs from the modification  induced by extra dimensions
in several aspects\cite{Arkani-Hamed:1998rs}.   First, in extra dimension ADD  models\cite{Arkani-Hamed:1998rs}
  the  corrections to the potential from extra dimensions falls  off exponentially at large distances $r/R>1$,
  where $R$ is a compactification length scale,    while at   short  $r/R<<1$,
   the $r$ dependence  has the form $1/r^{n+1}$ where $n$ is an integer.  This is to be contrasted with
Eq.(\ref{RG}) where the  correction from ungravity
 has the $r$ dependence of the form $1/r^{2d_{\cal U} -1}$ both at short  as well as  at large
 distances, and further $d_{\cal U}$ can take on non-integer  values.  Further, for the case of extra dimensions the  constraint that the physics of the
solar system not be modified eliminates  $n=1$\cite{Arkani-Hamed:1998rs}, 
and one has modifications of the Newtonian potential for
$n=2$ of the form $1/r^3$.
For the case of warped extra dimension model\cite{Randall:1999ee,Arkani-Hamed:1999hk,Chung:2000rg}
 the correction to the gravitational potential can
interpolate between $n=1$ and $n=2$  for the case with
small warpingl\cite{Chung:2000rg}, i.e.,
 between the form $1/r^2$ and $1/r^3$.
However, both for the   warped and the unwarped dimension case the
analytic form of the correction to the potential is significantly   different from
the one in  ungravity. Thus it should be  feasible to distinguish between extra  dimension models
 including models with warped  dimensions from the ungravity correction to the gravitational potential.

We note that purely kinematical corrections to Newtonian potential have
been computed in   general relativity~\cite{Donoghue:1994dn}.
  The sign of this correction  as well as its $r$ dependence differs from the one computed
here. Further, the effective  $R_{G}$ in this correction  is
$R_{GR}=G(m_1+m_2)/c^2$ and is of size the  Planck length
or smaller.
Thus in the context  of submillimeter experiments
these corrections are negligible.  Finally it is  interesting to note  that  renormalization group
analyses of quantum gravity in 4 and  higher
dimensions\cite{Fischer:2006fz}
 show that the graviton propagator  near an ultra violet fixed point scales as
$ {\cal G} (p)\sim 1/p^{2(1-\eta/2)}$ where  $\eta =4-D$ near the fixed point with $D$  the
number of space-time
dimensions. This propagator  has resemblance to the one that appears  in Eq.(\ref{propagator}).
Of course  the typical length scale in quantum gravity is  the Planck length while the length scale
in ungravity can lie in the submillimeter  region and be accessible  to  experiment.

The interaction operator $\kappa_* \sqrt g  T^{\mu}_{;\mu}  O^{\cal U}/\Lambda_U^{d_{\cal U}-1}$
can also play a role in  high energy scattering, and its domain of validity is also constrained from
that consideration. Consider the
process $f\bar f \rightarrow $ scalar unparticle ($f$ is a fermion), which would give
a Feynman amplitude
$
{\cal M} =$  ${m \bar u(p_1)v(p_2)}$ $/{\mstar\ \lu^{\du -1}}$
where $m$ is the  mass of the fermion and $p_1,p_2$ are the incident momenta. Using the notation
and phase space calculation of \cite{Georgi:2007ek}, we find a cross section
\be
\sigma(f\bar f\rightarrow {\cal U} )= \frac{1}{4s} \left(\frac{m}{\mstar}\right)^2
\ \left(\frac{\sqrt{s}}{\lu}\right)^{2\du-2}\ A_d\, \ .
\label{sig}
\ee
(Restriction to the inclusive reaction enables us to probe dimensions $d_{\cal U}>2$ without
encountering the pole term $\sin(\pi \du)$~\cite{Georgi:2007ek}.)
Since the annihilation to the unparticle proceeds through a single partial wave ($s$-wave),
the cross section is bounded by unitarity, $\sigma < 16\pi/s$. From
Eqs.(\ref{adu}) and (\ref{sig}) this gives an upper bound on the energy for compatibility of the
unparticle effective lagrangian with unitarity~\cite{greiner}:
\be
\sqrt{s} < \frac{1}{R_*}\ \left(\sqrt{\frac{64\pi}{A_d}}\ \frac{\mpl}{m}\right)^{\pd}
\label{sqrts}
\ee
where we have expressed the unitarity constraint in terms of the quantity
$R_* \equiv (1/\lu)( M_{Pl}/\mstar)^{\pd}$ proportional to the quantity $R_G$ defined
in Eq.~(\ref{RG}). The present upper bound on $R_G$ (see Fig. (\ref{fig12})) can be rewritten in terms of
$R_*$: for the region of interest $2<\du<3,$
a convenient parameterization $R_*^{max} \simeq (0.5 + 1.75(\du -2))\times 10^{12}\;
{\rm GeV}^{-1}$ will suffice.

If the exchange of the scalar unparticle is to be  consistent with the present
Newton's Law experiments, yet have a chance of showing up in future experiments, $R_*$
must lie below $R_*^{max}$ but above (say) $0.1\ R_*^{max}.$
Inserting this in (\ref{sqrts}), we obtain the following result:
for the worst-case scenario, with the fermion being the top quark, unitarity is not violated up to
1.2 TeV for $2 < \du< 2.3$. (Above this energy, rescattering corrections are significant.)
For the other fermions, of course, the range of validity is  larger.
Even if we require compatibility with perturbative QCD for the light quarks
(including the $b$-quark), which is
a tighter constraint, it  allows $2<\du<2.2$ for $\sqrt(s)_{max} \simeq 1.2$ TeV. There are similar
bounds if $T^\mu_\mu$ is saturated with the gluon trace anomaly.
To sum up, we can maintain compatibility of scale invariance with both high and low energy constraints,
and simultaneously not rule out seeing corrections to
Newton's Law in the next generation of gravitational experiments.

Corrections to Coulomb's law can also be similarly computed if one assumes couplings  of a vector unparticle operator
 $O^{\cal U}_{\mu}$
to the conserved em current $J^{\mu}$ with an interaction of type
$ (e_*/{\Lambda_U^{d_u-1}})  J^{\mu} O^{\cal U}_{\mu}$,
where $d_{\cal U} \geq 2$.
An analysis similar to the above  gives  the following modified Coulomb  potential
 \beqn
V_C(r)= \frac{K e_1e_2 }{r}  \left[ 1
+  \left(\frac{R_{C}}{r}\right)^{2d_{\cal U}-2}\right ],\nonumber\\
R_{C}=  \frac{1}{\pi\Lambda_U} \left(\frac{|e_*|}{|e|}\right)^{\frac{1}{d_{\cal U}-1}}.
\left(
\frac{2} {\pi}
\times \frac{\Gamma (d_{\cal U}+ \frac{1}{2}) \Gamma (d_{\cal U}- \frac{1}{2})
}   { \Gamma (2d_{\cal U})}\right)^{\frac{1}{2d_{\cal U}-2}}
\eeqn
Coulomb's law is not tested beyond the Fermi scale. Setting $R_{C} < 10^{-13}$ cm,
$d_{\cal U}=2$, and keeping $\Lambda_{\cal U} = 10^{-11}$ GeV, one
finds the constraint $e^*/e<10^{-11}$.
Thus a sensitive probe  could unravel the effects  of unparticle exchange to
Coulomb's law  below such scales.

In summary, we  have investigated  the implications of a scenario where conformal invariance of the hidden
sector strictly holds  down to very low  energies. This requires  constraining the dimensionality of the
scalar unparticle operators which might couple
to the  Higgs field so that $d_{\cal U}>2$ in order not to
spoil the conformal invariance of the hidden sector. Under the assumption that a traceless
rank two unparticle  operator can couple  to  the stress  tensor, we have computed  corrections to the inverse
square law and find scale invariant power law corrections which can be  discriminated  from similar corrections from extra  dimension models.  We also find the corrections from the exchange of a
scalar operator (with $d_{\cal U}>2$)  which couples to the trace of the stress tensor.
These corrections are testable in future experiments on the submillimeter
probes of gravity.
We note in passing that the analysis of spin 2 operators  in the context of collider phenomenology
is discussed in~\cite{spin2}.
Corrections to Coulomb's law from the exchange of vector  unparticle  operators were
also computed.

Finally we remark that the fractional modifications of the inverse square law  was 
studied by Dvali\cite{Dvali:2006su} and was seen to lead to strong coupling
effects.  Dvali's discussion was premised on infrared modifications of gravity which dominate the
Einsteinian term at a scale $r >> r_c$ which leads to the strong
coupling referred to above. However, in our case, the modification of
gravity at large scales does not dominate the Einsteinian term. In
momentum space, the conformal propagator goes like $P^{(2d -4)}$,
which for $d>1$ is suppressed relative to the Einstein case, $P^{(-2)}$,
while the propagator considered in the Dvali analysis  behaves as 
$P^{(-2\alpha)}, ~ (\alpha < 1)$, which indeed dominates the
Einsteinian term. Thus our set up escapes the strong coupling effect encountered
in \cite{Dvali:2006su}.\\

We end with a note of caution, in that
a fully consistent formulation of  unparticles does not  exist and this feature carries over also to
ungravity.  Nonetheless, if unparticle stuff exists,  and one assumes strict conformal invariance of the
hidden sector, a new gravitational size force, ungravity, could generate power law modification of gravity,
and  the new effects  fall within the range of  future submillimeter tests of gravity.
Further, it is possible to distinguish between modifications of corrections due to extra  dimensions
and corrections from  ungravity effects. It should be  interesting to build explicit models of the hidden
sector where strict conformal invariance is realized while also realizing couplings via a connector
sector to the Standard Model fields of the type discussed here.   The strict conformal invariance of the hidden
sector required by our
model   is also suggestive of an AdS5 connection. However, such issues lie outside the scope of this
work. 
Note added: After this work was done, another work\cite{Deshpande:2007mf} in a similar spirit examined the correction to the
long range forces from couplings to the  baryon and lepton number currents and found
that such corrections are significantly constrained by data.

\noindent
{\it Acknowledgements}:
We would like to thank   Hooman Davoudiasl,  Charlie Hagedorn,
Daniel Litim, and  Y. Nakayama for helpful communications.
This work is supported in part by NSF grants
PHY-0244507 and  PHY-0456568.


\begin{thebibliography}{999}

\bibitem{Banks:1982gt}
 T.~Banks and A.~Zaks,
 Nucl.\ Phys.\  B  {\bf 206}, 23 (1982).

\bibitem{Georgi:2007ek}
 H.~Georgi,
 Phys.\ Rev.\ Lett.\  {\bf 98}, 221601 (2007);
  Phys.\ Lett.\  B {\bf 650}, 275 (2007).


\bibitem{Cheung:2007ue}
  K.~Cheung, W.~Y.~Keung and T.~C.~Yuan,
  Phys.\ Rev.\ Lett.\  {\bf 99}, 051803 (2007);
  Phys.\ Rev.\  D {\bf 76}, 055003 (2007)



\bibitem{Luo:2007bq}
 M.~Luo and G.~Zhu,
 arXiv:0704.3532 [hep-ph];
 C.~H.~Chen and C.~Q.~Geng,
 arXiv:0705.0689 [hep-ph];
 G.~J.~Ding and M.~L.~Yan,
 arXiv:0705.0794 [hep-ph];
 Y.~Liao,
 arXiv:0705.0837 [hep-ph];
 T.~M.~Aliev, A.~S.~Cornell and N.~Gaur,
 arXiv:0705.1326 [hep-ph];
 S.~Catterall and F.~Sannino,
 arXiv:0705.1664 [hep-lat];
 X.~Q.~Li and Z.~T.~Wei,
 arXiv:0705.1821 [hep-ph];
 M.~Duraisamy,
 arXiv:0705.2622 [hep-ph];
 C.~D.~Lu, W.~Wang and Y.~M.~Wang,
 arXiv:0705.2909 [hep-ph];
 M.~A.~Stephanov,
 arXiv:0705.3049 [hep-ph];
 N.~Greiner,
 arXiv:0705.3518 [hep-ph];
 H.~Davoudiasl,
 arXiv:0705.3636 [hep-ph].
 D.~Choudhury, D.~K.~Ghosh and Mamta,
 arXiv:0705.3637 [hep-ph];
 S.~L.~Chen and X.~G.~He,
 arXiv:0705.3946 [hep-ph];
 T.~M.~Aliev, A.~S.~Cornell and N.~Gaur,
 arXiv:0705.4542 [hep-ph];
 P.~Mathews and V.~Ravindran,
 arXiv:0705.4599 [hep-ph];
 R.~Foadi, M.~T.~Frandsen, T.~A.~Ryttov and F.~Sannino,
 arXiv:0706.1696 [hep-ph];
 M.~Bander, J.~L.~Feng, A.~Rajaraman and Y.~Shirman,
 arXiv:0706.2677 [hep-ph].
 T.~G.~Rizzo,
 arXiv:0706.3025 [hep-ph].

\bibitem{Foxetal}
 P.~J.~Fox, A.~Rajaraman and Y.~Shirman,
  Phys.\ Rev.\  D {\bf 76}, 075004 (2007).




\bibitem{nakayama}
 Y.~Nakayama,
 arXiv:0707.2451 [hep-ph].

\bibitem{mack}
G.~ Mack, Commun. Math. Phys. {\bf 55}, 1(1977).


\bibitem{Appelquist:1996dq}
 T.~Appelquist, J.~Terning and L.~C.~R.~Wijewardhana,
 Phys.\ Rev.\ Lett.\  {\bf 77}, 1214 (1996).


\bibitem{Kapner:2006si}
 D.~J.~Kapner et.al., 
 Phys.\ Rev.\ Lett.\  {\bf 98}, 021101 (2007);
Liang-Cheng et.al., 
Phys.\ Rev.\ Lett.\ {\bf 98}, 201101 (2007).


\bibitem{Adelberger:2006dh}
 E.~G.~Adelberger et.al., 
 Phys.\ Rev.\ Lett.\  {\bf 98}, 131104 (2007).

%

\bibitem{Arkani-Hamed:1998rs}
 N.~Arkani-Hamed, S.~Dimopoulos and G.~R.~Dvali,
 Phys.\ Lett.\  B {\bf 429}, 263 (1998);
 I.~Antoniadis, N.~Arkani-Hamed, S.~Dimopoulos and G.~R.~Dvali,
 Phys.\ Lett.\  B {\bf 436}, 257 (1998).

\bibitem{Randall:1999ee}
 L.~Randall and R.~Sundrum,
 Phys.\ Rev.\ Lett.\  {\bf 83}, 3370 (1999)
 [arXiv:hep-ph/9905221].

\bibitem{Arkani-Hamed:1999hk}
 N.~Arkani-Hamed, S.~Dimopoulos, G.~R.~Dvali and N.~Kaloper,
 Phys.\ Rev.\ Lett.\  {\bf 84}, 586 (2000)
 [arXiv:hep-th/9907209].

\bibitem{Chung:2000rg}
 D.~J.~H.~Chung, L.~L.~Everett and H.~Davoudiasl,
 Phys.\ Rev.\  D {\bf 64}, 065002 (2001).

\bibitem{Donoghue:1994dn}
 J.~F.~Donoghue,
 Phys.\ Rev.\  D {\bf 50}, 3874 (1994)
 [arXiv:gr-qc/9405057].

\bibitem{greiner} See N.~Greiner, Ref.~\cite{Luo:2007bq} for another application of
unitarity.


\bibitem{Fischer:2006fz}
 P.~Fischer and D.~F.~Litim,
 Phys.\ Lett.\  B {\bf 638}, 497 (2006);
 AIP Conf.\ Proc.\  {\bf 861}, 336 (2006).


\bibitem{spin2}
P.~Mathews and V.~Ravindran, Ref.~\cite{Luo:2007bq};
K.~Cheung, W.~Y.~Keung and T.~C.~Yuan,Ref.~\cite{Cheung:2007ue}.



\bibitem{Dvali:2006su}
  G.~Dvali,
  New J.\ Phys.\  {\bf 8}, 326 (2006)
  [arXiv:hep-th/0610013].

\bibitem{Deshpande:2007mf}
  N.~G.~Deshpande, S.~D.~H.~Hsu and J.~Jiang,
  arXiv:0708.2735 [hep-ph].




\end{thebibliography}
\end{document}